\documentstyle[eqsecnum,aps,twocolumn]{revtex}

\newtheorem{lemma}{Lemma}
\newtheorem{definition}[lemma]{Definition}
\newtheorem{theorem}[lemma]{Theorem}

\newtheorem{corollary}[lemma]{Corollary}

\newenvironment{proof}{{\bf Proof:}}{\hfill$\Box$}
\begin{document}
\draft
\preprint{HEP/123-qed}
\title{Anonymous Oblivious Transfer}
\author{J. M{\"u}ller-Quade and H. Imai}
\address{Imai Laboratory, Institute of Industrial Science, The University of 
Tokyo}
\date{December $3^{rd}$, 2000}
\maketitle

\begin{abstract}
In this short note we want to introduce {\em anonymous oblivious
transfer} a new cryptographic primitive which can be proven to be
strictly more powerful than oblivious transfer. We show that all
functions can be robustly realized by multi party protocols with {\em
anonymous oblivious transfer}. No assumption about possible collusions
of cheaters or disruptors have to be made.

Furthermore we shortly discuss how to realize {\em anonymous oblivious
transfer} with oblivious broadcast or by quantum cryptography. The
protocol of {\em anonymous oblivious transfer} was inspired by a
quantum protocol: the {\em anonymous quantum channel}.
\end{abstract}

\section{Introduction}
In~\cite{BeaGol89,GolLev90,CreGraTap95} multi party protocols with
oblivious transfer were presented which can tolerate a dishonest
majority. These protocols work with perfect security if all players
cooperate. But already one disruptor can abort the protocol without
being detected. The contribution of~\cite{ImaMue00Eurocrypt} were
protocols more robust against disruption. The idea was to replace two
party subprotocols which failed by multi party protocols. Then either
these protocols did work or a cheater could be identified.

Unfortunately replacing an oblivious transfer where the sender or the
receiver refuses to coopertate by a multi party protocol weakens the
security of the protocol. In~\cite{ImaMue00Eurocrypt} we can observe a
trade off between the size of a tolerable collusion of active cheaters
(including disruptors) and the size of a collusion of passive cheaters
unable to obtain secret data.

In this paper we present the new cryptographic primitive {\em
anonymous oblivious transfer} and prove that it is strictly more
powerful than oblivious transfer. With this primitive we can realize
multi party protocols which  work with
perfect security or a cheater can be identified unambigiously. As we
cannot expect higher robustness and security than that we claim that  
anonymous oblivious transfer is the most powerful cryptographic
primitive which can achieve unconditional security. We recently
learned about independent work
in this direction carried out by~\cite{FitGarMauOst00}.

\section{Multi Party Protocols}
In a multi party protocol a set $P$ of players wants to correctly
compute a function $f(a_1,\dots,a_n)$ which depends on secret inputs
of $n$ players. Some players might collude to cheat in the protocol as
to obtain information about secret inputs of the other players or
to modify the result of the computation.
Possible collusions of cheaters are modelled by {\em adversary structures}
\begin{definition}
An adversary structure is
a monotone set ${\cal A}\subseteq 2^P$, i.\,e., for subsets
$S'\subseteq S$ of $P$ the property $S\in {\cal A}$ implies $S' \in
{\cal
  A}$.
\end{definition}

We assume that one set $A\in{\cal A}$ of players collude to cheat in
the protocol. These players take all their action based on their
common knowledge.  

The main properties of a multi party protocol are:
{\footnotesize
\begin{enumerate}
\item A multi party protocol is said to be ${\cal A}$-{\em secure} if
no single collusion from $\cal A$ is able to obtain information about
the secret inputs of other participants which cannot be derived from
the result and the inputs of the colluding players.
\item A multi party protocol is ${\cal A}$-{\em partially correct} if no
possible collusion
  can let the protocol terminate with a wrong result.
\item A multi party protocol is called $\cal A$-{\em fair} if no
  collusion from $\cal A$ can reconstruct the result of the multi
  party computation earlier then all honest participants together. No
  collusion should be able to run off with the result.
\end{enumerate}
}
We will be more strict here and demand robustness even against
disruptors. 
{\footnotesize
\begin{enumerate}
\item[2'] A multi party protocol is ${\cal A}$-{\em correct} whenever no
  single collusion from $\cal A$ can abort the protocol, modify its
result, or
  take actions such that some player gets to know a secret value.
\end{enumerate}
}

A protocol is called $\cal A$-{\em robust} if it has all of the above
properties. Note that we will allow only one collusion to cheat, but
we think of every single player as being curious, i.\,e., even if he
is not in the collusion actually cheating he will eavesdrop all
information he can obtain without being detected cheating

With oblivious transfer all multi party protocols can be realized with
perfect security if all players are
cooperating~\cite{BeaGol89,GolLev90,CreGraTap95}. But a collusion of
players can abort the calculation, see next section.

\section{Impossibility Results}

In this section we show that oblivious transfer is not able to
implement all multi party protocols in the presence of cheaters which
can derivate arbitrarily from the protocol. Not even together with a
broadcast channel. Protocols offering perfect secrecy of the inputs
can be aborted by  a collusion of
players.

\begin{lemma}\label{MPNoGo}
  Let $P$ be a set of players for which each pair of players is
  connected by a secure and authenticated oblivious transfer channel
  and each player has access to a broadcast channel.
  Then $\cal A$-robust multi party computations are possible for all
  functions if and only if no two sets of $\cal A$ cover $P\setminus
  \{ P_i\}$ for a player $P_i\in P$ or $|P|=2$.
\end{lemma}

\begin{proof}
Let $A$ and $B$ be two possible collusions covering $P\setminus \{ P_i
\}$, then oblivious transfer cannot be implemented $\cal A$-robustly
between players of $A$ and players of $B$. Between any two players
Alice $\in A$ and Bob $\in B$ the oblivious transfer channel does not
work, but it is not obvious for the player $P_i$ who is refusing to
cooperate. The player $P_i$ must assist Alice and Bob. As no other
player can assist we are in the three party situation with an
oblivious transfer channel only between Alice and $P_i$ and Bob and
$P_i$. For each bit being transferred from Alice to Bob the player
$P_i$ knows either as much as Alice about this bit or he knows as much
as Bob. The players Alice and Bob cannot agree on a bit known to both
without $P_i$ knowing it, too. Hence oblivious transfer from Alice to
Bob becomes impossible without $P_i$ having to learn a secret of Alice
or a secret of Bob.
\end{proof}

\section{Multi Party Protocols}

In the multi party protocols of~\cite{CreGraTap95,ImaMue00Eurocrypt} a
collusion of disruptors can abort the protocol if an assumption about
possible collusions of disruptors is violated. We would like to have
cryptographic primitives where every time a conflict arises a cheater
can be identified. Two such primitives are {\em global bit commitment}
and {\em undeniable oblivious transfer}. We will show in the following
that these primitives, defined below, can realize the subprotocols
needed in~\cite{CreGraTap95,ImaMue00Eurocrypt} relative to no
assumptions about possible collusions.

\begin{definition}
A {\em global bit commitment (GBC)} binds a player to all other
players to the same bit in a way that this bit cannot be changed with
a non negligible probability unless the player colludes with all other
players.
\end{definition}

\begin{definition}
An {\em undeniable oblivious transfer (UOT) protocol} from a player
Alice $\in P$ to a player Bob $\in P$ allows Alice to generate a GBC
for a bit $b$ in a way that Bob learns the bit $b$ with probability
$1/2$ and Alice cannot know if Bob learned $b$.
\end{definition}

Now we introduce the notions used for the multi party protocols.

\begin{definition}
A {\em global bit commitment with Xor} ({\em GBCX}) to a bit $b$ is a
GBC to bits $b_{1L}$, $b_{2L},\dots,$ $b_{mL},$ $b_{1R},\dots,$
$b_{mR}$ such that for each $i$ $b_{iL}\oplus b_{iR}=b$.
\end{definition}

One important ability of these bit cimmitments with Xor is given in
the next result, which is taken from~\cite{CreGraTap95}, but see also
references therein.

\begin{theorem}\label{COPYworks}
GBCX allow zero knowledge proofs of linear
relations among several bits a player is committed to using
GBCX. Especially (in)equality of bits or a bit string being contained
in a linear code.

Furthermore GBCXs can be copied, as proofs may destroy a GBCX.
\end{theorem}

\begin{proof}
We will not state a full proof here as it can be found
in~\cite{CreGraTap95}. But we will restate the copying procedure as it
is an important subprotocol of all of the following protocols.

Suppose Alice is committed to Bob to a bit $b$ and wants two instances
of this commitment. Then Alice ceates $3m$ pairs of global bit
commitments such that each pair Xors to $b$. Then all other player, by
coin tossing, randomly partition these $3m$ pairs in three subsets of
$m$ pairs, thus obtaining three GBCX and ask Alice to prove the
equality of the first new BCX with her GBCX for $b$. This destroys the
old GBCX and one of the new GBCX, but an honest Alice can thereby
convince all players that the two remaining GBCX both stand for the
value $b$.
\end{proof}

The basic building block for multi party protocols
of~\cite{CreGraTap95} are distributed bit commitments, where each
player is committed to a share of a bit.

\begin{definition}
A {\em distributed bit commitment (DBC) of a user} Alice  $\in
P$ to a bit $b$ consists of $n$ GBCX one created by each player of $P$
such that only Alice knows how to open all of them and the Xor of all
values ot the GBCX equals $b$.

An {\em intermediate result DBC} consists of  $n$ GBCX  such that no
subset of players unequal $P$ can know how to open all of the GBCX.
\end{definition}

\begin{lemma}
With a protocol for generating GBCX and a broadcast channel one can
realize a DBC of a user.
\end{lemma}

\begin{proof}
Each player generates a GBCX and opens the commitment to Alice. In
case of a conflict the player opens his GBCX publicly. Then Alice
creates a GBCX such that the parity bit is the bit she wanted to
create a DBC for.  Only Alice knows how to open all commitments as she
created one herself.
\end{proof}

The intermediate result DBCs are automatically generated by the multi
party protocols for these we need the key protocol of~\cite{CreGraTap95}.

\begin{definition}
Given two players Alice and Bob where Alice is committed to bits
$b_0,b_1$ and Bob is committed to a bit $a$. Then a {\em committed
oblivious transfer} protocol ({\em COT}) is a protocol where Alice inputs her
knowledge about her two commitments and Bob will input his knowledge
about his commitment and the result will be that Bob is committed to
$b_a$.

In a {\em global committed oblivious transfer} protocol ({\em GCOT})
all players are convinced of the validity of the commitments, i.e.,
that indeed Bob is committed to $b_a$ after the protocol.
\end{definition}

For the next result we use one-out-of-two UOT, which is the usual
one-out-of-two OT, but the sender is (by GBCs) committed to the two
bits the receiver can choose from. The standard reduction from
one-out-of-two OT to OT can be used to turn UOT into one-out-of-two
UOT.

\begin{lemma}
With UOT and an authenticated broadcast channel one can realize GCOT. 
\end{lemma}

\begin{proof}
We will essentially restate the GCOT protocol of~\cite{CreGraTap95}
and see that with one-out-of-two UOT instead of one-out-of-two OT any
conflict results in the identification of a cheater.

{\bf GCOT}$(a_0,a_1)(b)$
{\small
\begin{enumerate}
\item All participants together choose one decodable $[m,k,d]$ linear
code $\cal C$ with $k>(1/2+2\sigma)m$ and $d>\epsilon n$ for positive
constants $\sigma,\epsilon$, efficiently
decoding $t$ errors.
\item Alice randomly picks $c_0,c_1\in{\cal C}$, committs to the bits
$c_0^i$ and $c_1^i$ ($i\in \{1,\dots,m\}$) of the code words, and
proves that the codewords fulfil the linear relations of $\cal C$.
\item Bob  randomly picks $I_0,I_1\subset \{1,\dots,M\}$, with
$|I_0|=|I_1| = \sigma m,$ $I_1\cap I_0 =\emptyset$ and sets
$b^i\leftarrow \overline b$ for $i\in I_0$ and $b^i \leftarrow b$ for
$i\not\in I_0$.
\item Alice runs ${\rm UOT}(c_0^i,c_1^i)(b^i)$ with Bob who gets $w^i$
for $i\in \{1,\dots,m\}$.
Bob tells $I=I_0\cup I_1$ to Alice who opens  $c_0^i,c_1^i$ for each
$i\in I$.
\item Bob checks that $w^i = c_{\overline b}^i$ for $i\in I_0$ and
$w^i = c_{b}^i$ for $i\in I_1$, sets $w^i\leftarrow c_b^i$, for $i\in
I_0$ and corrects $w$ using $\cal C$'s decoding algorithm, commits to
$w^i$ for $i\in \{1,\dots,m \}$, and proves that $w^1\dots w^m\in
{\cal C}$.
\item All players together randomly pick a subset $I_2\subset
\{1,\dots,m\}$ with $|I_2|=\sigma m$, $I_2\cap I=\emptyset$ and Alice
opens $c_0^i$ and $c_1^i$ for $i\in I_2$.
\item Bob proves that $w^i = c_b^i$ for $i\in I_2$.
\item Alice randomly picks and announces a privacy amplification
function $h:\{0,1\}^m\rightarrow \{0,1\}$ such that $a_0 = h(c_0)$ and
$a_1 = h(c_1)$ and proves $a_0= h(c_0^1,\dots,c_0^m)$ and $a_1=
h(c_1^1,\dots,c_1^m)$.
\item Bob sets $a\leftarrow h(w)$, commits to $a$ and proves $a =
h(w^1\dots,w^m)$.
\end{enumerate}
}

A conflict between Alice and Bob can only appear in connection with
step 4 or step 5. If these two steps would be performed honestly then
all other steps can be checked by all other players and it becomes
immediately clear who is cheating. In a conflict in connection with
step 4 or step 5 Bob claims that Alice sent something inconsistent
over the oblivious transfer channel or Alice accuses Bob to not have
committed to what he received.

In case of a conflict Alice opens all bits of $c_0, c_1$ to which she
is committed by the UOT also she opens her GBCX to these codewords,
if she is not able to do it or unveils non code words or other
inconsistent information she is detected
cheating. The bits of $c_0, c_1$ do not give away any secret as these
are random code words. If Alices information is correctly unveiled and
is consistent with all her past actions (proofs) then Bob was cheating
if he did complain. If it was Alice complaining Bob has to prove zero
knowledgly that the bit string $w$ he is committed to equals $c_0$ or
equals $c_1$ if he is able to convince all other players Alice is
detected cheating (conflicts appearing during the proofs can be
resolved easily as it is obvious for every player who is cheating). 
%...wirklich so...?
\end{proof}

One other important property of multi party protocols is {\em
fairness}.  A multi party protocol is called {\em fair} if no
collusion of players can reconstruct the result of the protocol
earlier than all honest players. This problem is solved in the
literature~\cite{Cle89,GolLev90} and will not be discussed here.

Hence we have everything to follow the protocols of~\cite{CreGraTap95}
robustly and in the following we need only to prove that a certain
cryptographic primitive can realize GBC (or GBCX) and UOT and we know
that it is capable of realizing all multi party protocols with perfect
security and robustness.

\begin{theorem}
Given a set of players $P$ such that every player can generate global
bit commitments and we have an undeniable oblivious transfer between
every pair of players. Then all
functions can be computed $2^P$-robustly by multi party protocols.
\end{theorem}

\begin{proof}
First we note that we do not need a broadcast channel as generating a
GBC and unveiling it can be viewed as broadcasting. We now sketch the
phases of a multi party protocol following~\cite{CreGraTap95}.  To
implement oblivious circuit evaluation to realize arbitrary functions
we have to show the existence of an AND and a NOT function on DBCs and
clearify how a protocol is initialized and how it is ended.

{\bf Initialization Phase}: All players have to agree on the function
to be computed as well as on the circuit $F$ to be used, they have to
agree on an adversary structure $\cal A$ such that the protocol will
be $\cal A$ robust and all players have to agree on the
security parameters used and on a code $\cal C$ for the GCOT protocol.

Then all players create DBCs to commit to their inputs.

{\bf Computing Phase}: The circuit is evaluated using AND and NOT
gates on the input DBCs.  

An AND on commitments can be realized by the following protocol: 
Alice is committed to $a$ and Bob is committed to $b$. Then Alice
chooses a random bit $a'$ and runs GCOT$(a', a'\oplus a)(b)$ with
Bob who gets $b'$. We have $a'\oplus b' = a\wedge b$ because for $b =
0$ we have $b' = a'$ and hence $a'\oplus b'=0$, for $b=1$ we get $b'=
a\oplus a'$ and $a'\oplus b'=a$.

To evaluate an AND on DBCs we observe that $(\bigoplus_{i=1}^n a_i)\wedge
(\bigoplus_{j=1}^n b_j) = \bigoplus_{i,j=1}^n (a_i\wedge b_j)$.
From this we can conclude that an AND operation on DBCs can be realized
by $n^2$ GPAND one for each pair of players and Xor operations for
each player.

To implement
the NOT gate one player is picked who must invert his ``share''. This
players generates a new GBCX and proves that it is unequal to the GBCX
he held before. Note that the GCOT within the AND protocol
has to work only in one direction between every pair of players.
Sometimes one needs several copies of a DBC. A DBC is copied by
copying the GBCX it consists of. A GBCX can be copied by copying all
its BCX with the procedure of Theorem~\ref{COPYworks}. 

{\bf Revelation Phase}:
The result of a computation is hidden in DBCs. These have to
be unveiled in a way to ensure the fairness of the
protocol. Following~\cite{CreGraTap95} we use the techniques
from~\cite{Cle89,GolLev90}
to gradually unveil the secret information such that no collusion can
run off with an advantage of more than a fraction of a bit.
Of course an $\widetilde{\cal A}$-secure protocol cannot be more than 
$\widetilde{\cal A}$-fair.
\end{proof}

\section{Anonymous Oblivious Transfer}

We next define {\em anonymous oblivious transfer}.

\begin{definition}
An {\em anonymous oblivious transfer} ({\em AOT}) protocol allows a
player Alice $\in P$ to send a bit string $b_1\dots b_m$ to a player
Bob $\in P$ such that Bob receives each bit of the bit string with
probability $1/2$ or he receives $\perp$ which indicates that he will
not learn this bit. Alice cannot know which bits Bob
received. Furthermore Bob does not know which player sent the bit
string.
\end{definition}

For the following we will need some subprotocols which can easily be
realized by AOT. To realize them we need a message authentication
function ${\rm Auth}(x,y)$ which outputs a string which authenticats
the message $x$ with the secret $y$, see~\cite{PfiWai92,ChaRoi90} for
an unconditional signature scheme based on such a function and
anonymous transfer.

\begin{lemma}\label{AuthBroadcast}
With AOT one can realize an
authenticated broadcast channel.
\end{lemma}

\begin{proof}
Every player sends $l$ times anonymously a random number to
Alice. Alice sends her message $m$ to every player together with ${\rm
Auth}(m,r)$ for all random numbers $r$ Alice received. \footnote{This
can be seen as ``signing'' the message~\cite{PfiWai92,ChaRoi90}.}  Then every
pair of players compares the message they received. Either they are all
the same and the protocol was successful or two different messages
show up (one might be the empty message). Now two cases can happen:
\begin{enumerate}
\item The second message is correctly authenticated, then we have a
high probability (depending on $l$) that the sender Alice was cheating
or
\item the second message is not correctly authenticated.
\end{enumerate}
In both cases we repeat the protocol until one of the following cases
holds:
\begin{enumerate}
\item The protocol was successful.
\item Alice is in conflict with all other players and has to leave the
protocol. 
\item The players complaining about Alice are always the same, then
these must be cheating as Alice cannot know who sent which random
number.
\item Enough different correctly authenticated messages are found such
that the probability that Alice is cheating is above a certain
threshold and she is expelled from the protocol.
\end{enumerate}
\end{proof}

\begin{lemma}\label{AnonymousBroadcast}
With AOT one can realize anonymous message transfer and an anonymous
broadcast channel which can fail only $n$ times or someone leaves the
protocol.
\end{lemma}

\begin{proof}
To send a message anonymously one has to encode the message with an
error correcting code to cope with the erasures of the AOT.

For an anonymous broadcast Alice sends her message $m$ anonymously to
a player $P_i$. This player broadcasts the message. If he broadcasts
something wrong Alice is in conflict with this player, complains about
him using the authenticated broadcast, and picks another player $P_j$
to start the procedure anew. Either the anonymous broadcast will eventually
be successfull or Alice will leave the protocol as she is in conflict
with all other players.
\end{proof}

\begin{corollary}\label{LaterIdentification}
With AOT one can realize the anonymous message transfer and anonymous
broadcast of Lemma~\ref{AnonymousBroadcast} in a way that the
anonymous sender can later identify himself.
\end{corollary}

\begin{proof}
For an anonymous broadcast with later identification Alice
authenticates her message $m$ with $n$ random numbers which she sends
anonymously to the players. Each player receives one random number.

Then she anonymously broadcasts the thus authenticated message
according to Lemma~\ref{AnonymousBroadcast}. No other player is later
able to impersonate Alice as only she knows the secret random numbers
of the honest players.
\end{proof}

With these protocols we can realize GBCX.

\begin{lemma}\label{GBCXworks}
With AOT one can realize GBCX.
\end{lemma}

\begin{proof}
We let all players create GBCX according to the protocol
of~\cite{CreGraTap95}, but anonymously, using AOT and anonymous
broadcast. Then after some time no new conflicts occur for $l$
anonymous GBCX of each player ($l$ is a security parameter which is
polynomial in $n$). If a player Alice was unable to create a GBCX we
will split the set of players in a way that one set contains all
honest players and the other sets contain only cheaters.
We explain this in more detail by the
two cases which can occur:
\begin{enumerate}
\item If Alice was honest then, as a cheater cannot distinguish
between the honest players after some time if the cheater keeps
complaining about Alice this cheater will be in conflict with all
honest players. Furthermore all honest players will know it.
Now we can seperate the set $P$ of players several subsets such that 
all players in each subset are in conflict with the same players.
Then we can be
sure that one of the sets contains all honest players and every honest
player knows it.
\item If Alice was dishonest then we will also seperate the set
$P$. Alice will be in one group with all players complaining about the
same players as Alice did (these are all honest players if Alice were
honest) all other players will be in the other sets. As Alice is a
cheater and hence in conflict with an honest player all players in her
group must be cheaters, too.
\end{enumerate}
\end{proof}

Note that the protocol to create GBCX for all players needs only
polynomial time in $n$, as only $n^2$ conflicts are possible.

After having realized GBCX we need to implement UOT.

\begin{lemma}
With AOT one can realize UOT.
\end{lemma}

\begin{proof}
Alice creates a GBCX following Lemma~\ref{GBCXworks} and Bob publishes
positions of two substrings of the strings Alice sent to him. One
substring where he knows all the bits and one substring where he knows
nothing. The substrings must have approximately the same length.

Alice publishes the bits of one of the substrings. Then Bob either
learnt nothing new or he knows the bit Alice is committed to. We have
realized UOT if we can show that no other player learns the bit Alice
is committed to by the information published by Alice, but this is
trivial as Alice sent different strings to different players.
\end{proof}

%We recently got to know the abstract of a talk at the rump session of
%Crypto `00~\cite{FitGarMauOst00}. There an {\em oblivious cast} protocol is
%introduced which is a three party protocol also able to overcome the
%problem of unresolvable conflicts as there is always a third player
%involved. 

\section{Realizing AOT}

\subsection{Quantum Protocols}
Anonymous oblivious transfer was inspired by a quantum
protocol~\cite{ImaMue00PRL}. But it cannot be realized by a quantum
protocol unless no two possible collusions cover the set $P$ of
players. 

The idea for the realization is to follow normal quantum multi party
protocols~\cite{ImaMue00QMPforIEICE} if not two sets covering $P\setminus
P_i$ are in conflict. In case of such a conflict the player $P_i$ is
not a disruptor or active cheater by assumption. This player can now
forward quantum information between the two sets which are in
conflict.  Quantum cryptography allows to keep the player $P_i$ from
eavesdropping the quantum data excluding what happened in
Lemma~\ref{MPNoGo}. As the player $P_i$ can forward all quantum
information in the same way and send quantum information himself this
realizes an {\em anonymous quantum channel}. Together with the results
of~\cite{ImaMue00QMPforIEICE} we get:

\begin{theorem}
Robust quantum multi party protocols for all functions are
possible if and only if no two possible collusions cover the set
$P$ of players.

These protocols become robust against a set of possible collusions
after termination which may contain one and only one complement of a
collusion tolerable during the execution of the protocol.
\end{theorem}

For a proof see~\cite{ImaMue00PRL}.

Especially a quantum channel can be more powerfel than oblivious
transfer (See Lemma~\ref{MPNoGo}).  For details please refer
to~\cite{ImaMue00QMPforIEICE,ImaMue00PRL}.

\subsection{Oblivios Broadcast}

%Noisy broadcast was, like OT, originally proposed to realize secure
%key exchange. In this section we will prove that the cryptographic
%power of noisy broadcast is equivalent to the power of AOT.

We can think of each player broadcasting weak signals. Signals which
can be received only with a certain probability which is independent
for all receiving players. In this subsection we want to show that
this primitive is equally powerful as AOT. 

\begin{definition}
An {\em oblivious broadcast} channel is a protocol where a player
inputs a bit string and every other player receives the output of an
oblivious transfer of this string and the erasures are independent for
the different players. 
\end{definition}

\begin{lemma}\label{GBCviaOB}
An authenticated oblivious broadcast can realize a GBC.
\end{lemma}

\begin{proof}
Alice sends, as a commitment, $k$ bit strings of length $m$ ($k$, $m$
are security parameters which are polynomial in $n$) with parity $b$.
Then the knowledge all other players have about $b$ is negligible in
$m$. Because the probability that a bit is received by at least one
player is $1-1/2^n$ and the probability that all players together have
knowledge about all $m$ is $(1-1/2^n)^m$ which is negligible in
$m$. If $k$ strings are sent the probability remains negligible as $k$
and $m$ are polynomial in $n$.

If Alice wanted to change the bit she committed to she has to change
$k$ bits. The probability that any single player does not detect this
change is negligible in $k$.
\end{proof}

\begin{lemma}
An authenticated oblivious broadcast can realize UOT.
\end{lemma}

\begin{proof}
Alice creates a GBC and Bob publishes positions of two substrings of
the strings Alice sent over the oblivious broadcast. One substring
where he knows all the bits and one substring where he knows
nothing. The substrings must have approximately the same length.

Alice publishes the bits of one of the substrings. Then Bob either
learnt nothing new or he knows the bit Alice is committed to. We have
realized UOT if we can show that no other player learns the bit Alice
is committed to by the information published by Alice. But as the
substrings published are statistically independent of what the other
players received this information just changes the probability of
receiving a bit for each player. This change of probability can be
coped with an suitable choice of the security parameters used in
Lemma~\ref{GBCviaOB}.
\end{proof}

%...Note thyat if one uses one satellite sending weak signals ... one
%has to provide an authenticated broadcast channel as well...

\section{Main Result}

Summarizing all of the above we can state:

\begin{theorem}
The primitive of anonymous oblivious transfer is cryptographically
strictly more powerful than oblivious transfer. It can realize all
multi party protocols with a security and robustness which is
independent from assumptions about possible collusions of cheaters or
disruptors. 

Anonymous oblivious transfer can be realized by an authenticated oblivious
broadcast channel or by a quantum protocol if no two possible
collusions cover the set of players.
\end{theorem}

\section{Future Work}

An interesting question is if a noisy broadcast channel is of the same
power as AOT. This seems to be clear for small sets of players, but if
the number of players grow large the difference between the error
probabilities possible for different collusions becomes large, too.
If all players collude against the sender the probability of error is
much lower as if all players collude against the receiver. To kope
with this problem will be an interesting direction of future research.

There probably are many other primitives of a cryptographic power
equivalent to AOT. This has to be investigated to maybe find
primitives which can be realized more easily or more efficiently
(compare~\cite{FitGarMauOst00}).

%%%%%%%%%%%%%%%%%%%%%%%%%%%%%%%%%%%%%%%%%%%%%%%%%%%%%%%%%%%%
%
% The Literature
%
%%%%%%%%%%%%%%%%%%%%%%%%%%%%%%%%%%%%%%%%%%%%%%%%%%%%%%%%%%%%


\begin{thebibliography}{1}

\bibitem{ImaMue00Eurocrypt}
Anonymous.
\newblock Multi party protocols with oblivious transfer.
\newblock The manuscript can be obtained via the authors of this paper, October
  2000.

\bibitem{BeaGol89}
D.~Beaver and S.~Goldwasser.
\newblock Multiparty computations with faulty majority.
\newblock In {\em Proceedings of the 30$^{th}$ FOCS}, pages 468--473. IEEE,
  1989.

\bibitem{ChaRoi90}
D.~Chaum and S.~Roijakkers.
\newblock Unconditionally secure digital signatures.
\newblock In A.~J. Menezes and S.~A. Vanstone, editors, {\em Advances in
  Cryptology: Crypto '90}, volume 537 of {\em LNCS}, pages 206--215.
  Springer-Verlag, Berlin, 1990.

\bibitem{Cle89}
R.~Cleve.
\newblock Controlled gradual disclosure schemes for random bits and their
  applications.
\newblock In {\em Advances in Cryptology: Crypto '89}, pages 573--590, Berlin,
  1989. Springer-Verlag.

\bibitem{Cre97}
C.~Crepeau.
\newblock Efficient cryptographic protocols based on noisy channels.
\newblock In {\em Advances in Cryptography: Eurocrypt 97}, Lecture Notes in
  Computer Science. Springer Verlag, 1997.

\bibitem{CreGraTap95}
C.~Crepeau, J.~van~de Graaf, and A.~Tapp.
\newblock Committed oblivious transfer and private multi-party computations.
\newblock In {\em Advances in Cryptology: Proceedings of Crypto `95}, pages
  110--123. Springer, 1995.

\bibitem{FitGarMauOst00}
M.~Fitzi, J.~Garay, U.~Maurer, and R.~Ostrovsky.
\newblock Oblivious cast and multi party protocols.
\newblock Rump session of Crypto 2000, August 2000.

\bibitem{GolLev90}
S.~Goldwasser and L.~Levin.
\newblock Fair computation of general functions in presence of immoral
  majority.
\newblock In A.~J. Menezes and S.~A. Vanstone, editors, {\em Advances in
  Cryptology: Crypto '90}, volume 537 of {\em LNCS}, pages 77--93.
  Springer-Verlag, Berlin, 1990.

\bibitem{ImaMue00PRL}
J.~M{\"u}ller-Quade and H.~Imai.
\newblock Quantum cryptographic three party protocols.
\newblock Los Alamos preprint quant-ph/0010111, October 2000.

\bibitem{ImaMue00QMPforIEICE}
J.~M{\"u}ller-Quade and H.~Imai.
\newblock Temporary assumptions for quantum multi party protocols.
\newblock Technical Report of ISEC~11 Technical Meeting, Tokyo, The paper can
  be obtained via the authors of this paper, 2000.

\bibitem{PfiWai92}
B.~Pfitzmann and A.~Waidner.
\newblock Unconditional byzantine agreement for any number of faulty
  processors.
\newblock In {\em Proc. STACS'92}, volume 577 of {\em LNCS}, pages 339--350.
  Springer-Verlag, Berlin, 1992.
\newblock This paper generalizes the result of~\cite{ChaRoi90}.

\end{thebibliography}
\end{document}